\documentclass[onecolumn]{article}
\usepackage{graphicx}
\usepackage{amsmath}
\begin{document}
\title{Persistent currents in carbon nanotubes based rings}
%
%Unit\'e PCPM, Universit\'e Catholique de Louvain, B\^atiment Boltzmann, 
%Place Croix du Sud 1, B-1348 Louvain-la-Neuve, Belgium \\
\author{Sylvain Latil\thanks{Present address: Unit\'e PCPM, 
Universit\'e Catholique de 
Louvain, Louvain-la-Neuve, Belgium} \\ 
GDPC, Universit\'e Montpellier II, France 
\and 
Stephan Roche \\
Commissariat \`a l'Energie Atomique, \\DSM/DRFMC/SPSMS, Grenoble, France 
\and
Angel Rubio\\
Departamento de F{\'\i}sica de Materiales, \\
Facultad de Ciencias Qu{\'\i}micas, UPV/EHU, \\ and
Donostia International Physics Center (DIPC),\\ 
San Sebasti\'an/Donostia, Spain}
\maketitle
\begin{abstract}
Persistent currents in rings constructed from carbon nanotubes
are investigated theoretically. After studying the contribution of  
 finite temperature or quenched disorder on covalent rings, the complexity 
due to the bundle packing is addressed. The case of
 interacting nanotori and self-interacting coiled nanotubes are
 analyzed in details in relation with experiments.
\end{abstract}
%\pacs{PACS numbers: 73.23Ra, 73.22.-f} 
%%%%%%%%%%%%%%%%%%%%%%%%%%%%%%%%%%%%%%%%%%%%%%%%%%%%%%%%%%%%%%%%%%%%%%%%%%%%%%%
%
\hyphenation{nano-tube nano-tubes syn-the-sized ex-pe-ri-ment-al-ly}
\renewcommand{\thesection}{\Roman{section}}
\section{Introduction and background}
Carbon nanotubes are micrometer-long hollow cylinders 
with nanometer scale radius, and electronic 
properties strongly depending on their geometrical helicity 
\cite{Saito_BOOK}.
As promising tools for building up nanoscale electronic devices 
\cite{Tans_TRANSISTOR,Martel}, 
nanotubes also allow to challenge the well established common 
theories of mesoscopic physics. Recently, many works of 
quantum transport in these systems have revealed
puzzling properties resulting from the mixing between their nanometer 
and micrometer combined length 
scales\cite{Todorov_LPM,Roche_AB,Frank_MWNTbal,Triozon,Suzuura}.

Nanotubes are very particular in the sense that
they lie in between small molecular systems such as benzene-type ring
molecules, and mesoscopic systems such as metallic or
semiconducting wires. In the former entities,  basic electronic 
conduction properties are entirely monitored by HOMO-LUMO gap and
discrete features of the  spectrum, whereas quantum wires may manifest, apart
from ballistic  transport, band conduction allowing quantum
interference phenomena whenever full coherence of the wave function is 
maintained over a reasonable scale. 

In metallic single wall nanotubes, at the charge neutrality point 
there exists some evidence of a Luttinger liquid behavior, namely 
that electron-electron interactions are 
strong enough to deviate the electronic status from Fermi 
liquid\cite{Bockrath_LL}. Besides,
 a recent work shows that the tube-tube interaction
 between different nanotube layers is able to induce a transition from 
the Luttinger liquid to a strongly correlated Fermi system or a Fermi 
liquid\cite{EggerGogolin}. However, other results on scanning tunneling 
spectroscopy\cite{Lieber} are fully interpreted in terms of the independent 
electrons model indicating the importance of screening effects. 
    
Rings of single-walled carbon nanotubes have been 
synthesized experimental\-ly \cite{Liu_CIRCLES,Martel_CIRCLES}.  
Depending on the circumference
length of the ring, a transition from n-type semiconducting
to metallic behavior was observed experimentally
 (from $\sim 60$ nm to $\sim 1$ mm)\cite{Watanabe} 
and further discussed theoretically\cite{Cubernati}.
Magneto-transport experiments on rings have also been 
performed\cite{Shea_MRneg}, manifesting negative magneto-resistance and
 weak electron-electron
interactions in the low temperature regime. However, no clear
Aharonov-Bohm effect was found, in opposition with the properties of
multiwalled nanotubes 
(MWNTs) case\cite{Schonenberger_MWNT_AB,Fujiwara}. 

An applied magnetic flux $\Phi$ is known to induce ring currents in 
molecules \cite{London}, and persistent currents (PC) in mesoscopic 
rings \cite{Bouchiat_PC}. In fact, every thermodynamic functions of the system 
are $\Phi_0$-periodic functions of the flux, where $\Phi_0=h/e$ is the flux 
quantum. Investigation of PC then yields valuable informations 
to understand both quantum coherence and dephasing 
rates\cite{Trivedi,Kravtsov}. 
These non dissipative currents are intimately related 
with the nature of eigenfunctions of isolated rings and their 
flux sensitivity\cite{Bouchiat_PC}. 
The manifestation of $\Phi_{0}$ periodic oscillations of 
magneto-resistance or persistent
currents is a generic feature that depends on the degree of
disorder and employed averaging procedure \cite{Bouchiat_PC,Trivedi}. Let us 
introduce a dimensionless flux $\phi=\Phi/\Phi_0$. For  non-interacting 
electrons, the total current at zero 
temperature is calculated by the following expression :
\begin{equation}
\label{persistent_current}
I_{\text{pc}}=-\sum_n\frac{\partial \epsilon_{n}}{\partial\Phi}
=-\frac{1}{\Phi_0}\sum_n\frac{\partial \epsilon_{n}}{\partial \phi}
\end{equation}
with the summation running up to the last occupied energy levels. 
At finite temperature $T$, all the states participate formally to 
the total free energy, and
\begin{equation}
\label{kiki}
I_{\text{pc}}(T)=-\frac{1}{\Phi_0}\sum_n\frac{\partial f_{n}}{\partial \phi}
\end{equation}
with 
\begin{equation}
f_n=-k_BT\ln\left[1+\exp 
\left(\frac{\epsilon_{F}-\epsilon_{n}}{k_BT}\right)\right]
\end{equation}
where $\epsilon_{F}$ is the Fermi level, and $k_B$ the Boltzmann constant.
In ballistic mesoscopic rings, perfect agreement between 
theory and experiments has been reported \cite{Mailly}, 
differently to diffusive systems (i.e. the mean free path $l_{e}$ becomes 
smaller than the ring circumference $L_{\text{ring}}$), for which the 
discrepancy between the theoretical and the experimental values of PC
($I_{\text{pc}}^{theor.}\sim 10^{-2}I_{\text{pc}}^{exp.}$) remains an open 
controversy, with electron-electron interactions\cite{Ambegaokar,Giamarchi}, 
confinement effects\cite{Appel} 
or non-equilibrium phenomenon\cite{Kravtsov} as possible explanations. 
Persistent currents in a Luttinger liquid has been analyzed in\cite{LLPC}.

The goal of the present work is to establish to which extent 
the intrinsic features of carbon
nanotubes based ring geometries (helicities of individual tubes, tube-tube
interaction,\ldots) are reflected on the persistent current patterns in the 
coherent regime, neglecting electron-electron interactions.

%
%%%%%%%%%%%%%%%%%%%%%%%%%%%%%%%%%%%%%%%%%%%%%%%%%%%%%%%%%%%%%%%%%%%%%%%%%%%%%%%
%   
\section{Persistent currents in a simple carbon torus}
Different works have been published recently on the magnetic properties of 
simple carbon tori \cite{Lin,Marganska}, both are formally based on a 
tight-binding (TB) approach. 
In this paper, the same model is used. It consists in a finite length $(n,m)$ 
nanotube (containing $N$ primitive cells) curled up 
onto a torus. Such torus will be named like $(n,m)\times N$. 
If the circumference $L_{\text{ring}}$ of the torus is long enough, the 
effect of the curvature on the electronic properties is weak and 
the topologically equivalent system of 
a straight nanotube with periodic boundary conditions keeps the 
essential physics. 
The band structure of this straight tube is calculated with the zone 
folding technique \cite{Saito_BOOK}, keeping only one electron per site and 
assuming a constant hopping $\gamma_0$ between the first neighboring sites.
The resulting flux dependent hamiltonian is
\begin{equation}
\label{equ:H}
H = -\gamma_{0} \sum_{i, j(i)} | i \rangle \langle j |
.\exp(i \varphi_{ij})
\end{equation} 
\begin{equation}
\label{equ:phase}
\varphi_{ij}=2\pi.\left(\frac{z_j - z_i}{L_{\text{ring}}}\right).\phi
\end{equation}
where $| i \rangle$ and $| j \rangle$ are the $\pi_\bot$-orbitals 
located on site, 
whose positions along tube axis are called $z_i$ and $z_j$, and 
$\phi=\Phi/\Phi_0$ is the dimensionless magnetic flux.
Since the ring contains $N$ primitive cells, the 
first Brillouin zone is sampled for a number $N$ of $k$-points, 
equally separated by $\Delta k=2\pi/L_{\text{ring}}$, 
and the linearization of the band structure near the Fermi points implies 
that the energy level spacing is $\Delta_E=h.v_F/L_{\text{ring}}$.
In this case, the magnetic flux $\Phi$ through the ring 
acts just as a dephasing rate on the 
electronic levels, i.e. the sampling of $k$ points in the first Brillouin 
zone is shifted about $\delta k=\Delta k.\phi$.

As reported by Lin and Chuu \cite{Lin}, there are two kinds of metallic 
carbon tori. If the Fermi moment is zero, or if the number $N$ of cells is a 
multiple of 3, then the torus possesses a vanishing HOMO-LUMO gap at 
zero field (type I). If the Fermi moment is non zero, and the 
number of cells is not a multiple of 3, then the torus is sorted onto 
type II, with a non vanishing gap at zero field. 
On the other hand, a semiconducting carbon torus do not 
exhibit any flux dependent effect \cite{Note_SC_tori}.

The physical origin of the persistent currents is the cyclic boundary condition
of the electronic system, leading to a sampling of $k$-points 
in the Brillouin zone. 
\mbox{FIG.1} (left) shows this sampling in the vicinity of the 
Fermi energy, and the resulting discrete spectrum of electronic levels. 
The application of a magnetic flux dephases the sampling and moves 
linearly these levels 
(Zeeman effect) because each $k$-point is related to a magnetic 
momentum. The linear $\phi$-dependence of 
the individual levels is shown in FIG.1 (right).
The resulting persistent current $I_{\text{pc}}(\phi)$, obtain by summing the 
contribution of each occupied level, 
is a set of affine functions of the magnetic flux, as shown on FIG.2. 
For the type I with $T=0$, it is:
\begin{eqnarray}
I_{\text{pc}}(\phi) &=& I_{0}(-1-2\phi),  \quad\forall \phi\;\in\; [ -1/2,0[\\ 
I_{\text{pc}}(\phi) &=& I_{0}(1-2\phi),  \quad\forall \phi\;\in\;]0,1/2]
\end{eqnarray}
and for the type II:
\begin{eqnarray}
I_{\text{pc}}(\phi) &=& I_{0}(-1-2\phi),  \quad\forall \phi\;\in\;[-1/2,-1/3[\\
I_{\text{pc}}(\phi) &=& -2I_{0}\phi,  \quad\forall \phi\;\in\;]-1/3,1/3[\\
I_{\text{pc}}(\phi) &=& I_{0}(1-2\phi),  \quad\forall \phi\;\in\;]1/3,1/2]
\end{eqnarray}
with the value $I_0$ given by
\begin{equation}
\label{equ:I_0}
I_0=2N_{c}ev_{F}/L_{\text{ring}}
\end{equation}
with $N_c$ the number of available channels at Fermi level; $N_c=2$ 
for metallic nanotubes. Hence, the slope of the function $I_{\text{pc}}$ for 
metallic carbon nanotori is determined by its length only.
The reason why the function $I_{\text{pc}}$ is not analytical at the points 
$\phi=0$ and $\phi=\pm 1/3$ is a crossing of eigenenergies at Fermi 
level (on FIG.1 right), which 
provokes a sharp diamagnetic-paramagnetic transition. This work is 
mainly devoted to the first category of carbon nanotori. 

Since $I_{\text{pc}}(\phi)$ is a periodic and odd function, it can be written
down as a Fourier serial like : 
\begin{equation}
I_{\text{pc}}(\phi)=\sum_{n=0}^\infty b_n.\sin(2\pi n\phi)
\end{equation}
with $b_n$ the $n^{\text{th}}$ harmonic 
\begin{equation}
b_n = 2\int I_{\text{pc}}(\phi).\sin(2\pi n\phi)d\phi
\end{equation}
%
%
%%%%%%%%%%%%%%%%%%%%%%%%%%%%%%%%%%%%%%%%%%%%%%%%%%%%%%%%%%%%%%%%%%%%%%%%%%%%%%%
%
\section{Effect of Temperature and disorder}
In order to investigate the evolution of $I_{\text{pc}}$, with different  
physical parameters, such as static disorder or electronic temperature, 
the behavior of two typical currents is analyzed hereafter. 
The first one is the quadratic flux-averaged current 
\begin{equation}
J_{\text{quad}}
=\sqrt{
\int_{-1/2}^{+1/2}
I_{\text{pc}}^{2}(\phi)d\phi}
\end{equation}
and the second is the absolute value of $I_{\text{pc}}$ for the quarter of 
the quantum flux, which is related to the intensity of the first harmonic 
\begin{equation}
J_{1/4} = \left |
I_{\text{pc}}\left(\frac{1}{4}\right)
\right |
\end{equation}
We first analyze the effect of temperature on the intensity 
of PCs. Here, we still assume that the phase coherence length 
$L_\varphi$ is much larger than $L_{\text{ring}}$, so that the system remains 
fully described by the eigenfunctions of its hamiltonian. The electrons are 
independent and the effect of temperature is to induce a distribution 
of the occupancy around $\epsilon_F$, within $k_BT$. 
$I_{\text{pc}}(\phi)$ is given by (\ref{kiki}). 
As shown in the left part of FIG.3, the effect of $T$ is a  
smoothing of the sawtooth shape of the PCs, 
which became fully analytical functions. This means that the harmonics of 
higher order are the first to be affected by temperature.
On the right side of FIG.3 is plotted the evolution of the 
typical currents versus the temperature. 
The behavior is simple : below a critical temperature $T^*$ the
function $J_{\text{quad}}$ decreases slightly and $J_{1/4}$ is a constant.  
Above the transition, both are exponentially damped. 
The effect of $T$ on the PC of linear chains or 1D free 
electrons systems has been investigated by Cheung 
\emph{et al.}\cite{Gefen_Chain}, who demonstrated 
that for every shape of band structure, a temperature transition can be 
defined as:
\begin{equation}
T^*=\frac{\Delta_E}{2\pi^2}
\end{equation}
where $\Delta_E=h.v_F/L_{\text{ring}}$ is the level spacing at Fermi energy.
When $T>T^*$ the function $I_{\text{pc}}(\phi)$ is fairly approximated by its
first harmonic, and since this harmonic is the last to be affected, the 
typical currents decrease drastically after the transition.

It is worth to note that the behavior of such tubular 
system is exactly the same that a linear chain or 1D free electrons 
systems. This effect is due to the small number of channels available in a 
carbon nanotube ($N_c=2$). In fact, none effect of the radius of 
the nanotube happens.  In that 
sense, defect free metallic nanotubes behave like real one dimensional systems.

We turn now to the analysis of the effect of a quenched disorder 
on the persistent current, with a particular focus on 
the typical value of the averaged current, in order to compare the results 
with established theories. 
To discuss the effect of conduction mechanism on persistent
currents, static disorder is simulated by a random modulation of the 
on-site energies \cite{Anderson}. Particular attention is paid to the 
behavior of PC close to the transition between ballistic and more 
localized conduction regimes. Since the application of this static disorder 
breaks the translational symmetry, 
the whole structure (and no longer one unit cell only) is conserved to solve
the eigenproblem. As a generic case of metallic tubes the $(6,6)$ armchair 
nanotube (radius $R\simeq 2.58$ nm) is considered. 
The length of the tube is 75 unit cells, implying a perimeter of about 
$L_{\text{ring}}\simeq 18.4$ nm, and a single particle level spacing at Fermi 
energy $\Delta_E\simeq0.205$ eV. 
Randomly fluctuating on-site energies are added to the former 
$\phi$-dependent TB hamiltonian (\ref{equ:H}) like
\begin{equation}
H = \sum_i \varepsilon(i)  | i \rangle \langle i |
-\gamma_{0} \sum_{i, j(i)} | i \rangle \langle j |
.\exp(i \varphi_{ij})
\end{equation} 
where $\varphi_{ij}$ is given by (\ref{equ:phase}), and $\varepsilon(i)$ is 
the randomly modulated on-site energy of the $i^{\text{th}}$ atom. 
The range of the on-site energies fluctuation is $[-W/2,W/2]$, where $W$ is
the disorder strength.
This modulation enables to tune an important 
physical transport length scale in nanotubes, namely
the electronic mean free path which given by (for the armchair $(n,n)$ 
nanotubes \cite{Todorov_LPM,Roche_AB})
\begin{equation}
\label{equ:lpm}
l_{e} \simeq  6\sqrt{3} n a_{\small\rm cc}\left(\frac{\gamma_{0}}{W}\right)^2
\end{equation}
where $a_{\small\rm cc}=1.42$ {\AA} 
is the distance between two carbon atoms of the honeycomb 
lattice.
Since this formula derives from the Fermi golden rule, it should be 
valid in the limit of weak scattering and for Fermi energies close to the 
charge neutrality point. 
An important feature, which is not common in mesoscopic wires, is 
that, in first approximation, the mean free path scales with the 
diameter.

In mesoscopic systems, the quantity $l_{e}$ discriminates between three
different physical conduction regimes: the ballistic motion for
$l_{e} > L_{\text{ring}}$, the diffusive regime $l_{e}<L_{\text{ring}}$ 
 and the localized regime whenever 
$\xi\simeq N_{c}l_{e} < L_{\text{ring}}$ \cite{Imry_BOOK}. 
A single level of energy $\epsilon_n$ carry a current $I_{n}^{2}(\phi)$.
In the diffusive regimes of mesoscopic systems,
the typical current carried by this single level $J_{\text{typ}}^{(n)}$ 
reads \cite{Gefen_Chain,Cheung_PRL,Mont_PC} 
\begin{equation}
J_{\text{typ}}^{(n)}=\sqrt{\langle\overline{I_{n}^{2}(\Phi)}
\rangle_{W,L_{\text{ring}}}}
\sim \frac{\sqrt{\Delta E_{c}}}{\Phi_{0}}
\end{equation}
where $\langle ...\rangle$ denotes the average value over disorder strength
and system length, and $\overline{I_{n}^{2}(\Phi)}$ 
is the flux-average current.
The typical total current 
has been found to be in order of 
\begin{equation}
J_{\text{typ}}=\sum_{n}J^{(n)}_{\text{typ}}
\sim\frac{E_{c}}{\Phi_{0}} = \frac{ev_{F}l_{e}}{L_{\text{ring}}^{2}}
\end{equation}
where $E_{c}=\hbar D/L_{\text{ring}}^{2}=\hbar v_{F}l_{e}/L_{\text{ring}}^{2}$ 
is the Thouless energy (diffusivity constant is $D$). 
Differently, the effect of disorder on persistent 
current remains a complex non-elucidated problem in the 
ballistic limit\cite{Altland_THOULESS}.
 
To extract the effect of disorder on persistent 
currents, we thus follow their $W$-dependence. A first observation 
is that both behaviors of $J_{\text{quad}}$ and $J_{1/4}$ are 
reminiscent to the temperature dependent patterns. We plot the 
$W$-dependence of the two typical currents on FIG.4.a), for a number of 
random disorder configurations, and the resulting average. 
The behavior of $J_{1/4}$ is particularly illustrative since 
its flux dependence is mainly dominated by the first harmonic. 
We also plot on FIG.4.b) the dependence of  $J_{\text{quad}}$ and $J_{1/4}$
upon the equivalent mean free path $l_e(W)$, that is given by equation (18).
Both are exponentially damped as soon as 
$2l_{e} < L_{\text{ring}}$
which corresponds to 
$W\geq 1.83\times\gamma_{0}$.

For smaller disorder, behaviors become more 
difficult to interpret within the framework of conventional theory. 
For a diffusive regime, one would expect an averaged persistent current 
following $\sim ev_{F}l_{e}/L^{2}_{\text{ring}}$ so reduced by a factor of 
$\sim l_{e}/{L_{\text{ring}}}$ with respect to the ballistic case. 
This would yield a $(\gamma_{0}/W)^{2}$ dependence of the amplitude with 
disorder strength. However, from our results, the damping of $J_{1/4}$ and  
$J_{\text{quad}}$ are much smaller especially for $l_{e} < L_{\text{ring}}$.
It is indeed rather difficult to extract the $W$-dependence in the ballistic 
regime, since all harmonics respond differently to a given disorder strength. 
Actually, as long as $nL_{\text{ring}}<l_{e}$, 
the harmonic of rank $n$ remains 
nearly unaffected by disorder, whereas harmonics with 
$nL_{\text{ring}}>2l_{e}$ are exponentially damped.

In conclusion except from strong disorder cases, persistent currents in 
covalent rings of carbon nanotubes are weakly sensitive to disorder and 
their amplitude remain close to that of the clean ballistic case.
%
%
%%%%%%%%%%%%%%%%%%%%%%%%%%%%%%%%%%%%%%%%%%%%%%%%%%%%%%%%%%%%%%%%%%%%%%%%%%%%%%%
%   
\section{From a simple torus to a bundle}
The systems studied in the preceding sections are useful to understand the 
physical phenomena in carbon nanotubes, but seem far to be a reliable 
representation for describing the full complexity of nanotube based rings. 
The bonding nature of the closing of the ring of SWNTs has been suggested
either to be covalent\cite{Liu_CIRCLES} 
or of Van der Waals type\cite{Martel_CIRCLES}. 
In any case, the synthesized rings are constituted by SWNTs closely packed in 
bundles. They contains from tens to hundreds of individuals nanotubes, whose 
interaction are likely to control electronic pathways along the ring. 

The study of these complex system is difficult due to the large number of 
involved carbon atoms. First, the bundle packing effect will be scrutinized 
(drawn on FIG.5.a). Our goal is to understand how relevant is the Van der Waals
interaction between rings and how it affects the persistent current patterns.
Second, following some experimental observations
\cite{Martel_CIRCLES}, imperfect tori made with a curled nanotube 
(as shown in FIG.5.b) will be under consideration. 
These two models allow to envision the bundle effect on large ring systems.

The study of persistent currents, when torus-torus interaction is turn on, 
is done by constructing a model of bundle. It consists in sticking two of 
these objects (namely $A$ and $B$), and
allowing them to interact with an \emph{ad hoc} potential.
This tube-tube hopping, that we call $\gamma_1$, was optimized to 
reproduce correctly the electronic properties of 3D 
graphite\cite{Charlier_GRAPHITE} or MWNTs\cite{Meunier_Multi}. 
Such a tube-tube interaction between two tori is believed to affect the 
shape or the intensity of $I_{\text{pc}}(\phi)$ in various ways.
The hamiltonian is now
\begin{equation}
H= H_A + H_B - \gamma_1 
\sum_{<i,p>} \Big( | i \rangle\langle p | + | p \rangle\langle i | \Big) 
.\exp(i\varphi_{ij})
\end{equation}
$H_A$ and $H_B$ are the hamiltonians (\ref{equ:H}) 
of the non-interacting rings, and 
$| i \rangle$ (resp. $| p \rangle$) is a \mbox{$p_\bot$-orbital} 
of the $A$ (resp. $B$) 
torus. The phase $\varphi_{ij}$ is given by (\ref{equ:phase}) and 
the secondary hopping integral is
\begin{equation}
\label{equ:gamma_1}
\gamma_1=V_{\text{int}}\exp\left(\frac{d-\delta}{l}\right)
\end{equation}
where $d$ denotes the relative distance between the two sites. 
We take the value $\gamma_0=2.9$ eV, which allows the TB model to 
fit experimental data \cite{hopping}. 
The other parameters \mbox{are :} $V_{\text{int}}=0.36$ eV, 
$\delta =3.34$ \mbox{\AA} {and} $l= 0.45$ \AA. 
To simplify the comparison with precedent sections, $\gamma_0$ is kept as 
the energy unit. 

It is important to note that the two subsystems are forced to have the same 
circumference, hence the field corresponding to $\Phi=\Phi_0$ is the same in 
the two cases. If there are multiple correspondences between magnetic 
flux and field, and the $\Phi_0$-periodicity is ill defined. 
As such situation may happens, we should avoid it to extract the effect 
of the tube-tube interaction.
The choice of the subsystems is then restricted to equivalent 
helicities (e.g. two armchair tubes, two zigzag tubes, two ($3n$,$2n$), etc.). 
The distance between the tube walls is fixed to $\simeq 3.4$ \AA.

Our calculations show first that a semiconducting tube has no effect 
on the persistent current of a metallic one. 
We can imagine the distribution of states of the interacting tori as two 
quasi-continua under interaction. The resulting quasi-continuum, which
is the set containing all the eigenstates of the total system, is a mixing of
the eigenstates of each non-interacting subsystem. However the mixing rate is 
proportional to $\delta(\epsilon_i-\epsilon_j)$, where $\epsilon_i$ and 
$\epsilon_j$ are the two considered eigenvalues of each free torus. Hence, 
none coupling within the gap of the semiconducting torus happen. 
More, the persistent current in the metallic torus is carried by the levels 
around $\epsilon_F$, where the semiconducting tube does not have any state, 
then the persistent current of the complex system is equal to the persistent 
current of the free metallic torus.

The interesting cases arise whenever two metallic tori are interacting. 
The studied bundle is made with two (7,7) tori containing 84 primitive cells 
(they are type I tori, with a degenerated Fermi level).
By varying the relative orientations of the two tubes before curling them, 
and since the interaction $\gamma_1$ depends upon the geometry of the two 
subsystems, different cases of interaction schemes are explored.
As shown on FIG.6 a) and b), the tube-tube interaction induces a 
HOMO-LUMO gap at zero flux, and breaks level degeneracies. 
The flux-dependent evolution of the discrete levels shows that 
several crossing between eigenstates are driven by increasing magnetic flux, 
until the pattern returns to its original configuration when $\Phi=\Phi_{0}$. 
Since the value of the discrete energy levels changes from one orientation 
to another, the different level crossings 
induce modifications of the shape of $I_{\text{pc}}(\phi)$. However the slope 
of the currents, which depends on the circumference and the Fermi velocity 
only, is unaffected. On FIG.7 is shown the function $I_{\text{pc}}(\phi)$
for the two former cases and for the free tori is shown.
Hence in all cases, the tube-tube interaction will reduce the typical 
persistent current when compared to the isolated torus. 

Obviously, the precedent system is far to be a generic case, since 
the chosen torus is one of the highest symmetry. 
More, each subsystem is just the mirror image of the other one. 
The effect of incommensurability is then addressed considering two
different metallic SWNTs, with similar diameter (the $(7,7)$ and $(11,2)$
have same radii and the length cell of the latter is exactly seven times 
the length cell of the former). After checking that the magnetic properties of 
the $(7,7)\times 84$ torus are exactly the same than the $(11,2)\times 12$, 
the study of the bundle composed by one 
$(7,7)$ torus containing 84 periods and one $(11,2)$ torus containing 12 
periods is made. In this case even though the ring is by definition
$L_{\text{ring}}$-periodic ($L_{\text{ring}}$ = 2.07 nm here), 
the disorder-free bundle becomes quite different from the
previous ones since there is now incommensurability at short scale. 
The resulting discrete levels are plotted as function of flux on FIG.6.c). 
Close to the charge neutrality, the evolution of the levels is very similar 
to those of the single torus. Then, in the incommensurate case, the tube tube
interaction is not able to open an HOMO-LUMO gap, and the persistent current 
(also shown in FIG.7) is roughly equivalent to twice the one of the free 
torus.

Consequently, in this case, the tube-tube interaction could act as a 
static disorder, with a modulation strength 
$W=\left\langle\gamma_1\right\rangle\sim\gamma_0/10$.  
As discussed in the previous part, this range of disorder induces no  
relevant perturbations on the persistent currents, and explain why the PC of 
non-commensurate tori are not affected by tube-tube interactions. 

Finally, a last check is done. The FIG.6.d) shows the discrete levels of the 
$(11,2)+(2,11)$ structure. The splitting of levels is present, and is of the 
same order of magnitude than that the $(7,7)+(7,7)$ case. This means that 
the physical reason of absence of perturbation of PC is an effect of 
short range incommensurability and is not intrinsic to the underlying 
chirality.
%
%%%%%%%%%%%%%%%%%%%%%%%%%%%%%%%%%%%%%%%%%%%%%%%%%%%%%%%%%%%%%%%%%%%%%%%%%%%%%%%
%
\section{Coiled nanotubes}
In this part we are concerned with coiled nanotubes. These structures are 
imperfectly closed nanotori, which are rolled up with finite length nanotubes. 
The two ends of 
the nanotube are sticked through a Van der Waals interaction, like shown 
on FIG.5.b).The circumference of the curled system $L_{\text{ring}}$ 
is necessary
shorter than the length of the uncurled structure, otherwise sticking is
impossible. The length of the sticking area 
is noted $L_{\text{stick}}$, while the distance between the two tube's ends 
is $d_{\text{stick}}$. Of course, these three parameters can vary within 
a given bundle, and depending on their relative values, different pattern of 
$I_{\text{pc}}(\phi)$ will result.

If we keep restricting the working space to the $|p_\bot\rangle$ orbitals, the 
electronic hamiltonian takes the following expression
\begin{equation}
H= H_0 - \gamma_1 
\sum_{<i,p>} \Big( | i \rangle\langle p | + | p \rangle\langle i | \Big) 
\end{equation}
with $H_0$ the hamiltonian of the uncurled structure, and
$\gamma_1$ defined by equation (\ref{equ:gamma_1}). Now the sum runs for 
the orbitals $| i \rangle$ (resp. $| p \rangle$) located on the right 
(resp. left) extremity of the nanotube. 
Since the second part is clearly less 
important than the first it can be seen as a perturbation.
This means that the most relevant part of the hamiltonian 
describe an uncurled finite length nanotube. Such a kind of structures has 
been intensively investigated during the last years 
\cite{Liu_SizeFX,Rochefort_SizeFX,Rubio}. They exhibit a 
``particle-in-a-box'' behavior which imply an important interplay 
between the total length $L_{\text{ring}}+L_{\text{stick}}$, 
chirality and HOMO-LUMO gap properties: the gap is zero when the Fermi 
moment is non-zero, and the number of primitive cells within the structure 
is $3q+1$ (i.e. corresponding to the type I tori), else the gap
is non-zero (corresponding to the type II tori). The 
principal difference with the torus is that the rotational symmetry is lost, 
and the resulting electronic pathways around the ring are completely 
driven by the inter-tube hopping at the edges. The probability of hopping 
from one extremity to the other one will then depend on the geometry 
of the eigenstates around $\epsilon_F$, then upon the
length and the chirality of the used nanotube. 

Calculations were done on the (5,5) nanotube, with open ends. 
The parameter $d_{\text{stick}}$ monitors the interlayer hopping value, 
since $\gamma_1$ depends on the distance (equation(22)). 
As shown on FIG.8 (top), the results show that the intensity of 
$I_{\text{pc}}(\phi)$ is weakly dependent on the parameter $d_{\text{stick}}$, 
whereas the shape is not affected. The dependence on the sticking length 
$L_{\text{stick}}$ is drawn on FIG.8 (bottom), 
keeping $L_{\text{ring}}$ constant, showing the resonances according 
to the ``particle-in-a-box'' model. The shape is strongly dependent on the 
length --or the number of unit cells-- of the curled tube : adding one unit 
cell could switch the response from diamagnetic to paramagnetic, and if the 
length is resonant (the gap is minimal), the function $I_{\text{pc}}(\phi)$ is 
quasi ${\Phi_0}/{2}$-periodic. In all cases, the intensity of PCs is one 
order of magnitude lower than in the ideal case.
%
%%%%%%%%%%%%%%%%%%%%%%%%%%%%%%%%%%%%%%%%%%%%%%%%%%%%%%%%%%%%%%%%%%%%%%%%%%%%%%%
\section{Slater-Koster approach}
Until now, our study of the electronic properties of nanotori was limited to
$|p_\bot\rangle$ orbitals only. This approach is equivalent to the zone 
folding technique. We know that this model present some discrepancies compared
to more refined calculations. More precisely, the curvature needed to roll-up 
the sheet onto a tube, and the resulting $\sigma-\pi$ mixing are responsible 
of a gap opening in the band structure of all the non-armchair metallic 
nanotubes, and for a light shifting of the Fermi moment $k_F$ of armchair 
nanotubes, that keep a real metallic behavior. Although these modifications
are not spectacular, they affect directly the states just around the Fermi 
level. Since these states are carrying the persistent current, we need to 
study $I_{\text{pc}}(\phi)$ for nanotori described by a more accurate approach 
than the zone folding technique.

Calculations were done with an orthogonal tight-binding hamiltonian with 
Slater-Koster description of the hopping integrals. The parameters used are 
those of Louie and Tom\'anek\cite{Tomanek_Louie_SK}, 
but trials with Charlier, Gonze and Miche\-naud\cite{Charlier_Gonze_SK} 
parameters give the same results. The gap of metallic nanotubes is estimated 
$\simeq 0.106$ eV that is very large compared to the levels separation 
for zone folding method $\Delta E=2\pi v_{F}/L$. At the scale of this energy
$\Delta E$, the ``metallic'' nanotubes behave exactly like insulators : 
the magnetically induced current is small, as shown on FIG.9 for 
(9,0) based nanotorus.

However, the armchair nanotubes preserve band crossings at 
Fermi level, which means that armchair based nanotori (and only them) 
will still exhibit persistent currents. However, 
the Fermi moment $k_F$ is no more located at 2/3 of Brillouin zone, hence 
the concept of type I and II nanotori (with crossing of levels when $\Phi=0$, 
or not) looses its meaning. All the nanotori possess a non vanishing 
HOMO-LUMO gap for zero flux, and the crossing of levels does not happen 
exactly at the value $\Phi=\pm \frac{2}{3}\Phi_0$. As a consequence, 
there are more level crossings at charge neutrality point within the 
$[-\Phi_0/2, \Phi_0/2]$, as shown also on FIG.9, 
leading to a stronger weight of
the higher harmonics in the Fourier spectrum. Compared to the zone folding 
results, we can then expect an increased sensitivity of the induced current,
with temperature or disorder.
%
%%%%%%%%%%%%%%%%%%%%%%%%%%%%%%%%%%%%%%%%%%%%%%%%%%%%%%%%%%%%%%%%%%%%%%%%%%%%%%%
%
\section{conclusion and perspectives}
In this work, a study of persistent current in nanotube based rings has 
been presented. It was first point out that disorder and temperature 
diminish the persistent currents in nanotori in a similar fashion as it 
does in 1D systems, assuming a Fermi energy at the charge neutrality point. 
Further, it has been established that tube-tube interaction does not 
affect the shape and intensity of $I_{\text{pc}}(\Phi)$, except in the 
rare cases of commensurable systems. 

On the other hand, the necessity of intertube hopping to convey electrons 
along a closed path was shown to weakly 
damp the magnitude of typical PCs. More, if the 
curvature induced $\sigma-\pi$ 
mixing is taken onto account in the model, then the persistent currents are
completely screened, except for the armchair based nanotori.

This suggests that such quantity may fluctuate significantly from one 
bundle to another and that it should be strongly reduced in regards 
to the specific pattern found in pure tori\cite{Liu_coloss_MU}.If, for some reasons (doping, etc\ldots) the Fermi level is sufficiently 
shifted away from the charge neutrality point, then a linear dispersion 
relation is predicted for non-armchair metallic tubes. 
Hence in this case, the persistent current may be as strong as the one of 
armchair based nanotori.

This work is supported by European Community through the COMELCAN 
network (HPRN-CT-2000-00128). Pr. R. Saito and
 Pr. J.~C. Charlier are acknowledged for valuable discussions. 
%
%%%%%%%%%%%%%%%%%%%%%%%%%%%%%%%%%%%%%%%%%%%%%%%%%%%%%%%%%%%%%%%%%%%%%%%%%%%%%%%
%
\bibliographystyle{prsty}
%\bibliography{reference}

\begin{thebibliography}{10}

\bibitem{Saito_BOOK}
R. Saito, G. Dresselhaus, and M.~S. Dresselhaus, {\em Physical properties of
  carbon nanotubes} (Imperial College Press, London, 1998).

\bibitem{Tans_TRANSISTOR}
S. Tans, A. Verschueren, and C. Dekker, Nature {\bf 393},  49  (1998),
A. Bachtold, P. Hadley, T. Nakanishi and C. Dekker, 
Science {\bf 294},  1317  (2001).

\bibitem{Martel}
R. Martel, T. Schmidt, H.R. Shea, T. Hertel, Ph. Avouris, 
Appl. Phys. Lett. {\bf 73},  2447  (1998),
F. Leonard and J. Tersoff, Phys. Rev. Lett. {\bf 88},  258302  (2002),
S. Heinze, J. Tersoff, R. Martel, V. Derycke, J. Appenzeller and Ph. Avouris
, Phys. Rev. Lett. {\bf 89},  106801  (2002),
J. Appenzeller, J. Knoch, V. Derycke, R. Martel, S. Wind, Ph. Avouris , 
Phys. Rev. Lett. {\bf 89},
 126801(2002).

\bibitem{Todorov_LPM}
C. White and T. Todorov, Nature {\bf 393},  240  (1998),
T. Ando, Semicond. Si. Technol. {\bf 15},  R13  (2000).

\bibitem{Roche_AB}
S. Roche, G. Dresselhaus, M. Dresselhaus, and R. Saito, Phys. Rev. B {\bf 62},
  16092  (2000),
S. Roche and R. Saito, Phys. Rev. Lett. {\bf 87}, 246803 (2001).

\bibitem{Frank_MWNTbal}
P. Poncharal, C. Berger, Y. Yi, Z.L. Wang, and Walt A.
 de~Heer, J. Phys. Chem. {\bf 106},  12104 (2002).

\bibitem{Triozon}
S. Roche, F. Triozon, A. Rubio, and D. Mayou, Phys. Lett. A {\bf 285},  94
  (2001),\\
{\it ibid}, Phys. Rev. B {\bf 64}, 121401 (2001).

\bibitem{Suzuura}
H. Suzuura and T. Ando, Mol. Cryst. Liq. Cryst. {\bf 340},  731  (2000).

\bibitem{Bockrath_LL}
M. Bockrath, D. H. Cobden, J. Lu, A. G. Rinzler, R. E. Smalley, L. Balents, 
and P. L. McEuen, Nature {\bf 397},  598  (1999).

\bibitem{EggerGogolin}
R. Egger, Phys. Rev. Lett. {\bf 83},  5547  (1999), 
E. Arrigoni, Phys. Rev. B {\bf 61}, 7909 (2001), 
R. Egger and A. Gogolin, Phys. Rev. Lett. {\bf 87},  66401  (2001).

\bibitem{Lieber}
T.~W. Odom, J.-L. Huang, P. Kim and C.~M. Lieber, J. Phys. Chem. B {\bf 104},
  2794  (2000).
\bibitem{Liu_CIRCLES}
J. Liu, H. Dai, J.H. Hafner, D.T. Colbert, R.E. Smalley, S.J. Tans, C. Dekker, Nature {\bf 385},  780  (1997).

\bibitem{Martel_CIRCLES}
R. Martel, H.~R. Shea, and P. Avouris, Nature {\bf 398}, 299 (1999),
{\it ibid}, J. Phys. Chem. B {\bf 103}, 7551 (1999).

\bibitem{Watanabe}
H. Watanabe, C. Manabe, T. Shigematsu, and M. Shimizu, Appl. Phys. Lett. {\bf
  78},  2928  (2001).

\bibitem{Cubernati}
G. Cubernati, J. Yi, and M. Porto, Appl. Phys. Lett. {\bf 81},  850  (2002).

\bibitem{Shea_MRneg}
H. Shea, R. Martel, and P. Avouris, Phys. Rev. Lett. {\bf 84},  4441  (2000).

\bibitem{Schonenberger_MWNT_AB}
C. Sch\"onenberger, A. Bachtold, C. Strunk, J.-P. Salvetat and L. Forró, Appl. Phys. A {\bf 69}, 
283(1999).

\bibitem{Fujiwara}
A. Fujiwara, K. Tomiyama, H. Suematsu, M. Yumura, K. Uchida, Phys. Rev. B {\bf 60},  13492  (1999).
\bibitem{London}
F. London, J. Phys. Rad. {\bf 8},  397  (1937).

\bibitem{Bouchiat_PC}
M. Buttiker, Y. Imry, R. Landauer, Phys. Lett. A {\bf 96}, 365 (1983), 
L.~P. L\'evy, G. Dolan, J. Dunsmuir, and H. Bouchiat, Phys. Rev. Lett. {\bf
  64},  2074  (1990).

\bibitem{Trivedi}
N. Trivedi and D.~A. Browne, Phys. Rev. B {\bf 38},  9581  (1988).

\bibitem{Kravtsov}
V.~E. Kravtsov and B.~L. Altshuler, Phys. Rev. Lett. {\bf 84},  3394  (2000).


\bibitem{Mailly}
D. Mailly, C. Chapelier, and A. Benoit, Phys. Rev. Lett. {\bf 70},  2020
  (1993).

\bibitem{Ambegaokar}
V. Ambegaokar and U. Eckern, Phys. Rev. Lett. {\bf 65},  381  (1990).

\bibitem{Giamarchi}
T. Giamarchi and B.~S. Shastry, Phys. Rev. Lett. {\bf 51},  10915  (1995).

\bibitem{Appel}
V.~M. Appel, G. Chiappe, and M.~J. S\'anchez, Phys. Rev. Lett. {\bf 85},  4152
  (2000).

\bibitem{LLPC}
D. Loss, Phys. Rev. Lett. {\bf 69}, 343 (1992). 

\bibitem{Lin}
M.~F. Lin and D.~S. Chuu, Phys. Rev. B {\bf 57},  6731  (1998).

\bibitem{Marganska}
M. Marg\'anska and M. Szopa, Acta Phys. Pol. B {\bf 32},  427  (2001).

\bibitem{Note_SC_tori}
In fact, there is a little modification of the energy, due to a magnetization 
$M$ induced by the magnetic flux $\Phi$ as $M \sim \chi_m.\Phi$, where 
$\chi_m$ is the magnetic susceptibility, equivalent to susceptibility 
of straight nanotube. 
However since this magnetisation do not 
exhibit any $\Phi/\Phi_0$ behavior, we
ignore it.

\bibitem{Gefen_Chain}
H.~F. Cheung, Y. Gefen, E.~K. Riedel, and W.~H. Shih, Phys. Rev. B {\bf 37},
  6050  (1988), there is an arror in the legend of figure (3). The letter (a)
  labelizes actually the even number and (b) the odd.

\bibitem{Anderson}
P.~W. Anderson, Phys. Rev. {\bf 109},  1492  (1958).

\bibitem{Imry_BOOK}
Y. Imry, {\em Introduction to mesoscopic physics} (Oxford University Press, New
  York, 1997).

\bibitem{Cheung_PRL}
H. Cheung, E.~K. Riedel, and Y. Gefen, Phys. Rev. Lett. {\bf 62},  587  (1989).

\bibitem{Mont_PC}
H. Bouchiat and G. Montambaux, J. Phys. France {\bf 50}, 2695 (1989), 
G. Montambaux, H. Bouchiat and R. Friesner, 
Phys. Rev. B. {\bf 42}, 7647 (1990).

\bibitem{Altland_THOULESS}
A. Altland, Y. Gefen, and G. Montambaux, Phys. Rev. Lett. {\bf 76}, 1130 (1996).

\bibitem{Charlier_GRAPHITE}
J.~C. Charlier, J.~P. Michenaud, and P. Lambin, Phys. Rev. B {\bf 46},  4540 (1992).

\bibitem{Meunier_Multi}
P. Lambin, V. Meunier, and A. Rubio, Phys. Rev. B {\bf 62},  5129(2000).

\bibitem{hopping}
G. Dresselhaus {\it et~al.}, {\em On the $\pi - \pi$ overlap energy in carbon
  nanotubes \emph{in} ``Science and Application of Nanotubes'', \emph{D.
  Tom\'anek and R. J. Enbody ed.}} (Kluwer Academics, New-York, 2000).

\bibitem{Liu_SizeFX}
L. Liu, C.~S. Jahanthi, H. Guo, and S.~Y. Wu, Phys. Rev. B {\bf 64},033414 (2000).

\bibitem{Rochefort_SizeFX}
A. Rochefort, D.~R. Salahub, and P. Avouris, J. Phys. Chem. B {\bf 103},  641 (1999).

\bibitem{Rubio}
A. Rubio, D. Sánchez-Portal, E. Artacho, P. Ordejón, and J. M. Soler , Phys. Rev. Lett. {\bf 82},
3520(1999).

\bibitem{Tomanek_Louie_SK}
D. Tomanek and S.~G. Louie, Phys. Rev. B {\bf 37},  8327  (1988).

\bibitem{Charlier_Gonze_SK}
J.-C. Charlier, X. Gonze, and J.-P. Michenaud, Phys. Rev. B {\bf 43},  4579
  (1991),
P. Lambin, L. Philippe, J.~C. Charlier and J.~P. Michenaud, 
Comput. Mater. Sci. {\bf 2}, 350 (1994)

\bibitem{Liu_coloss_MU}
L. Liu, G. Guo, C. Jayanthi, and S. Wu, Phys. Rev. Lett. {\bf 88},  217206
  (2002),
S. Latil, S. Roche, A. Rubio and P. Lambin, comment submitted in 
Phys. Rev. Lett. 
\end{thebibliography}
%
%%%%%%%%%%%%%%%%%%%%%%%%%%%%%%%%%%%%%%%%%%%%%%%%%%%%%%%%%%%%%%%%%%%%%%%%%%%%%%%

%
%%%%%%%%%%%%%%%%%%%%%%%%%%%%%%%%%%%%%%%%%%%%%%%%%%%%%%%%%%%%%%%%%%%%%%%%%%%%%%%
%
%
\begin{figure}[t]
\label{FIG:EXPLI}
\includegraphics[width=\linewidth]{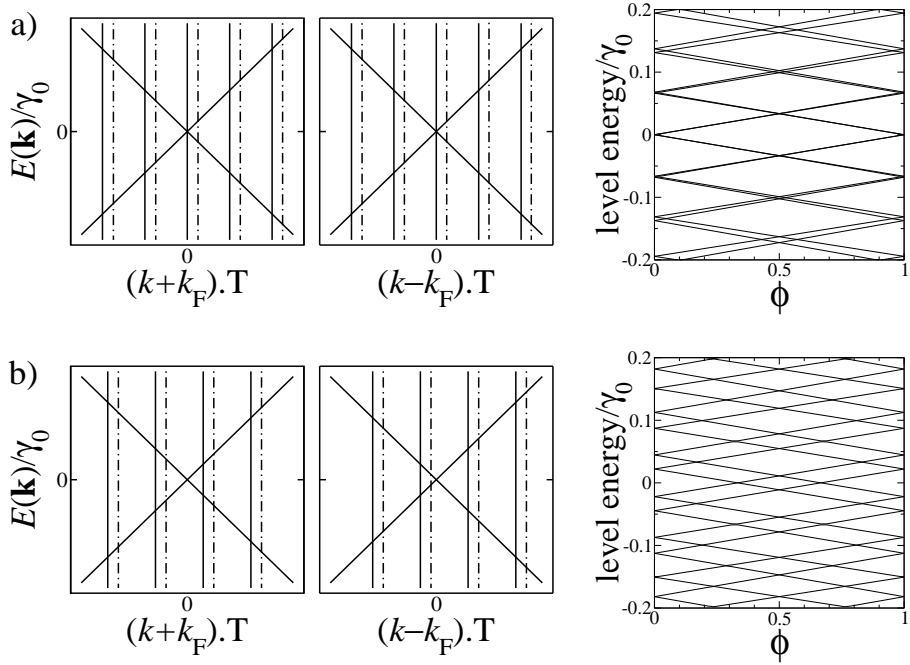}
\caption{The cyclic boundary conditions in the nanotorus produces a 
confinement, responsible of the sampling of the first Brillouin zone 
of the SWNT. At left a zoom of the vicinity of the crossing points is drawn, 
the vertical solid 
lines represent the sampling at zero flux, and the dot-dashed lines 
the sampling for a non-zero flux. At right, the 
resulting discrete electronic levels is plotted as a function of the 
dimensionless flux $\phi$. a) The HOMO-LUMO gap of the type I 
nanotorus vanishes at zero flux. Application of a magnetic field opens a 
gap by shifting the sampling of points. b) The pattern for the type II is more
involved since two crossing of levels occurs during the evolution of the 
flux.}
\end{figure}
\begin{figure}
\label{FIG:J_perm}
\includegraphics[width=\linewidth]{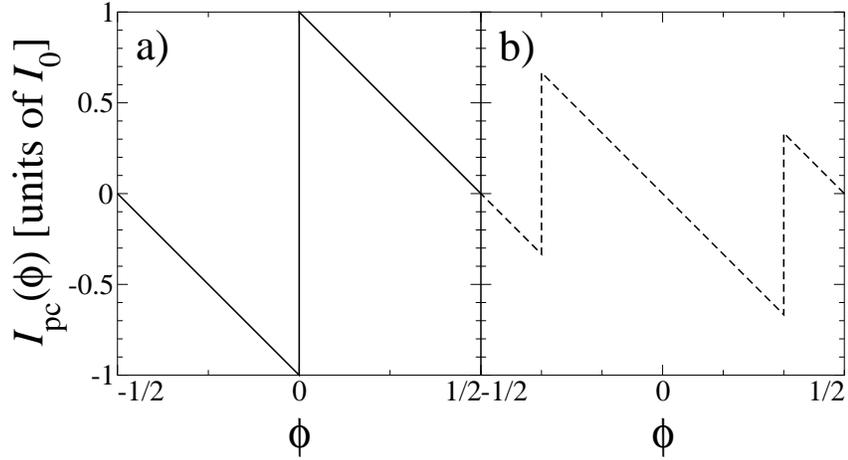}
\caption{$I_{\text{pc}}(\phi)$ for the two 
cases I (a) and II (b). The origin of the diamagnetic-paramagnetic transitions
are the crossing of levels seen on FIG.1. $I_0$ is given by (\ref{equ:I_0}).}
\end{figure}
\begin{figure}
\label{FIG:T}   
\includegraphics[width=\linewidth]{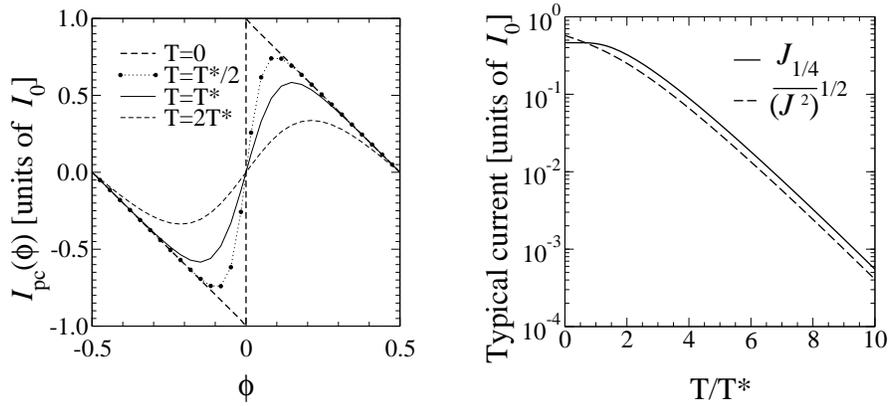}
\caption{Left : Temperature effect on the persistent $I_{\text{pc}}(\phi)$. 
Right : Semilog plot of the temperature dependent typical currents. 
As it happens for the linear chain of atoms {\cite{Gefen_Chain}}, 
these currents exhibit a transition temperature, namely $T^{*}$ 
(see text), from which the intensity decreases exponentially.}
\end{figure}
\begin{figure}
\includegraphics[width=\linewidth]{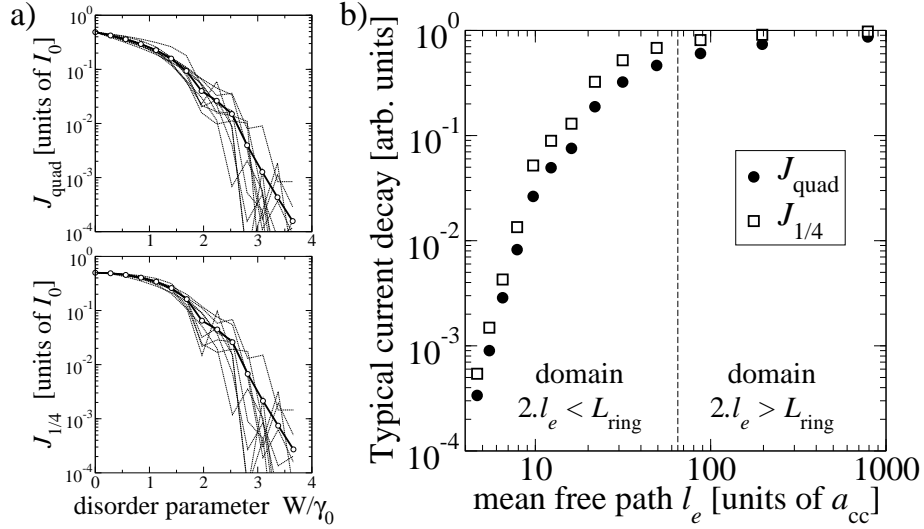}
\caption{a) Semilog plot of the typical currents $J_{\text{quad}}$ (up) 
and $J_{1/4}$ (bottom) versus the disorder strength. 
The dashed lines describe different disorder
configurations, their average is identified by the bold line with circles.
b) Log-log plot of the typical currents as function of mean free path $l_e$, 
given by equation (18). 
The transition from a ballistic (small $W$, long $l_e$) to a more 
localized (larger $W$, short $l_e$) is pointed out by a vertical line, 
at which $W$ yields $L_{\text{ring}}=2.l_e$.}
\end{figure}
\begin{figure}
\includegraphics[width=\linewidth]{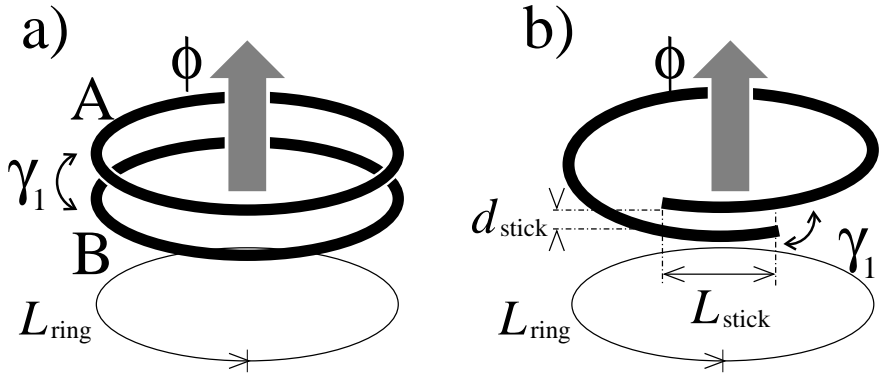}
\caption{a) A complex ring made of two interacting carbon nanotori. b) A 
coil shaped carbon nanotube.}
\end{figure}
\begin{figure}
\includegraphics[width=\linewidth]{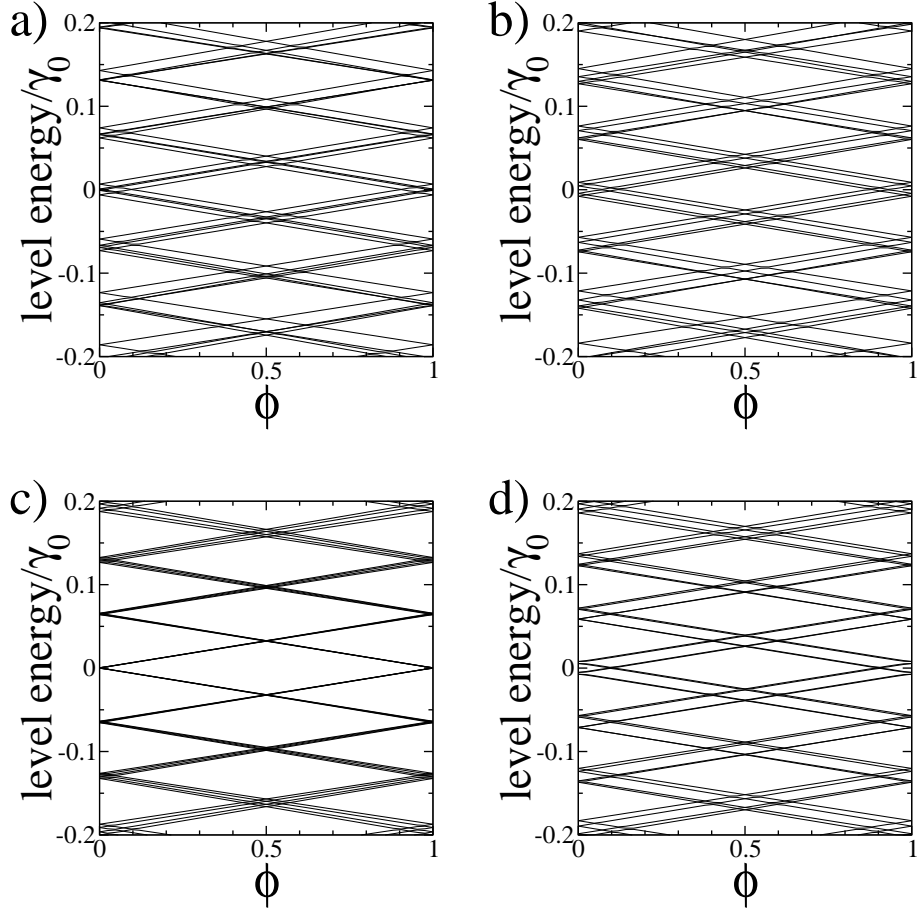}
\caption{Discrete levels of two interacting tori, as function of flux. 
a) Two $(7,7)\times84$ tori. The tube-tube interaction induces a lifting 
of degeneracies. b) Two $(7,7)\times84$ tori, with different 
orientation. c) A $(7,7)\times84 + (11,2)\times12$ non commensurate system. It
is fairly similar to the non-interacting case (see FIG.1).
d) A $(11,2)\times12 + (2,11)\times12$ commensurate system.}
\end{figure}
\begin{figure}
\includegraphics[width=\linewidth]{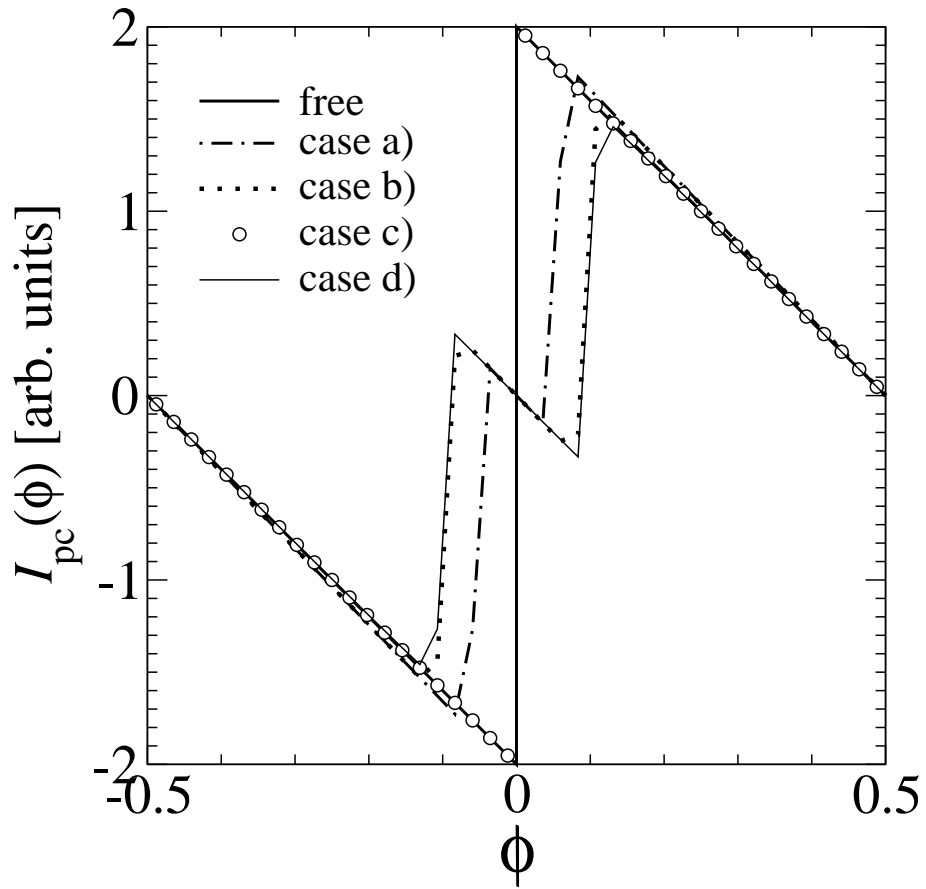}
\caption{Flux dependent persistent currents, for the same cases than
FIG.6. The shape $I_{\text{pc}}(\phi)$ for the case a), b) and 
d) is due to the crossing of levels at Fermi energy. The non-commensurate 
structure (case c) behaves similarly than the free tori. }
\end{figure}
\begin{figure}
\begin{center}
\includegraphics[width=8cm]{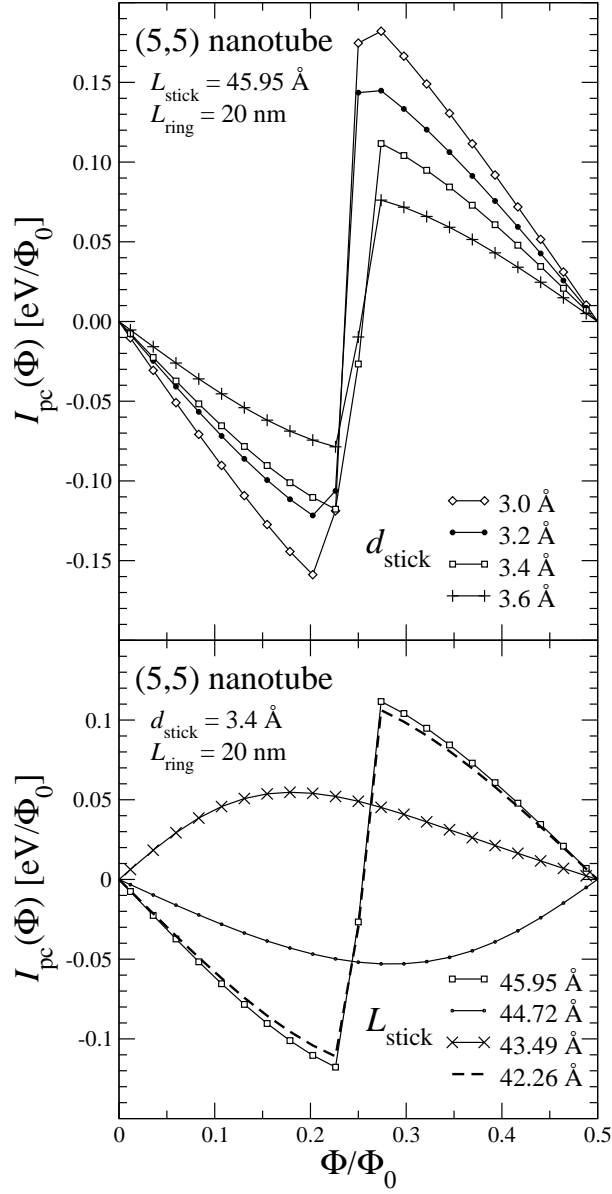}
\end{center}
\caption{Flux dependent persistent currents induced in coiled nanotubes.
 Top: Evolution with respect to $d_{\rm stick}$ for fixed $L_{\rm stick}$ and 
$L_{\rm ring}$. 
Bottom : Illustration of the effect of increasing $L_{\rm stick}$ 
while keeping  $L_{\rm ring}$ and $d_{\rm stick}$ unchanged.}
\end{figure}
\begin{figure}
\includegraphics[width=\linewidth]{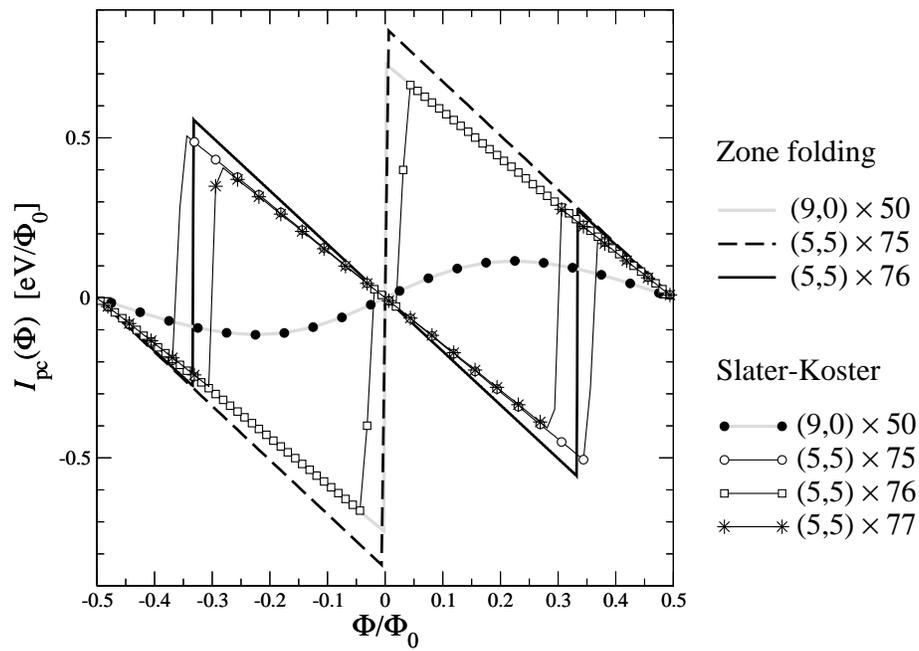}
\caption{Comparison of persistent currents obtained for the (9,0) nanotorus 
with the zone-folding method (thin lines) and the Slater-Koster approach 
(thick lines).}
\end{figure}
\end{document}